\documentclass[%
reprint,
showpacs,
amsmath,amssymb,
aip,
jap,
floatfix,
]{revtex4-1}

\usepackage{graphicx}
\usepackage{color}
\usepackage{dcolumn}
\usepackage{bm}

\newcommand*{\citen}[1]{%
  \begingroup
    \romannumeral-`\x 
    \setcitestyle{numbers}%
    \cite{#1}%
  \endgroup   
}

\def\eeq{\relax}
\def\beq#1#2\eeq{\begin{equation}\label{#1}#2\end{equation}}
\def\bal#1#2\eal{\begin{align}\label{#1}#2\end{align}}
\def\bse#1#2\ese{\begin{subequations}\label{#1}#2\end{subequations}}

\def\rev#1{\textcolor{blue}{#1}}	    
\def\reva#1{\textcolor{magenta}{#1}}	    

\begin{document}


\title{Elastic metasurfaces for splitting SV- and P-waves in elastic solids} 

\author{Xiaoshi Su}
 \email{xiaoshi.su@rutgers.edu}
 \affiliation{Mechanical and Aerospace Engineering, Rutgers University, Piscataway, NJ 08854}
\author{Zhaocheng Lu}%
\affiliation{Mechanical and Aerospace Engineering, Rutgers University, Piscataway, NJ 08854}
\author{Andrew N. Norris}%
\affiliation{Mechanical and Aerospace Engineering, Rutgers University, Piscataway, NJ 08854}

\date{\today}

\begin{abstract}
Although recent advances have made it possible to manipulate  electromagnetic and acoustic  wavefronts with sub-wavelength metasurface slabs, the design of elastodynamic counterparts remains challenging.  
We introduce a novel but simple design approach 
to control SV-waves in elastic solids. The proposed metasurface can be  fabricated by cutting an array of aligned parallel cracks in a solid such that the materials between the cracks act as plate-like waveguides in the background medium. The plate array is capable of modulating the phase change of SV-wave while keeping the phase of P-wave unchanged. An analytical model for SV-wave incidence is established to calculate the transmission coefficient and the transmitted phase through the plate-like waveguide explicitly. A complete $2\pi$ range of phase delay is achieved by selecting different thicknesses for the plates. An elastic metasurface for splitting SV- and P-waves is designed and demonstrated using full wave finite element (FEM) simulations. Two metasurfaces for focusing plane and cylindrical SV-waves are also presented. 
\end{abstract}

\maketitle


\section{\label{intro}Introduction}
Achieving full control of wave propagation with ultra-thin material slabs is of particular interest in engineering applications. In the past decade, the emerging area of metasurface research has made it possible to manipulate optical and electromagnetic waves in an almost arbitrary way by tuning the phase gradient at the sub-wavelength scale. \cite{Yu2011,Sun2012,Kildishev2013,Pfeiffer2013,Estakhri2016} This concept has also found applications in acoustic designs such as focal lenses, \cite{Li2012a,Wang2014,Li2014} anomalous reflection and refraction, \cite{Zhao2013,Zhao2013a,Dubois2017,Xie2014,Li2015,Li2016}, and generation of acoustic orbital angular momentum. \cite{Jiang2016,Shi2017} Elastic metasurfaces \cite{Zhu2016,Liu2017} are relatively unexplored; they present specific challenges due to the mode conversion at the material interface which makes the phase modulation more complicated. 

Recently, \citet{Zhu2016} designed and experimentally demonstrated a few metasurfaces for controlling mode converted and unconverted lamb waves in plates. However, their design approach involves mode conversion, as a result, the transmitted field contains unwanted wave types. We are interested in controlling different types of bulk waves individually without introducing others. We propose a new metasurface design to split SV- and P-waves in elastic solids into different propagation directions without involving mode conversion. Achieving full control of these types of waves may have applications in ultrasonics and nondestructive evaluations. The metasurfaces are designed by introducing an array of aligned parallel cracks in a bulk elastic medium. The materials separated by these cracks act as plate-like waveguides connecting two elastic half-spaces. For a normally incident SV-wave, the transverse vibration couples to the flexural waves in each plate without mode conversion. Then each of the plate-like waveguide serves as a phase modulator to achieve certain phase gradient for the transmitted wavefronts. The main idea  is that the phase speed of the SV-wave in the background material only depends on the material properties, while the flexural wave speed in the plate is sensitive to the thickness which makes it possible to achieve desired phase shift for the metasurface design. For instance, we can design a metasurface to change the propagation direction of the transmitted SV-wave by tuning the phase gradient according to the generalized Snell's law \cite{Yu2011}
\begin{equation}\label{general}
(\sin \theta_t-\sin \theta_i)k_T=d\phi/dy,
\end{equation}
where $k_T$ denotes the wavenumber of the SV-wave in the solid; $\phi$ is the phase of the transmitted wave; $\theta_i$ and $\theta_t$ are the incident and transmitted angle, respectively. The mechanism for normally incident P-wave is different in that the longitudinal wave speeds in the plate array are the same. Due to this feature, there is no phase difference in the transmitted wavefronts and therefore the P-wave travels along the incident direction.

In the applications proposed in this paper, the phase gradients are small so that the phase modulation is more critical than the amplitude modulation. In order to predict the transmitted phase accurately, we first establish an analytic model to calculate the transmission coefficient of the unit cell. Then we take advantage of this model to select the thicknesses of the plates for the metasurface designs. The same transmission problem has been considered by \citet{Su2016}, but the model was based on a thin plate assumption which is only valid for a low frequency range, i.e.~$kh\ll 1$. In this paper, we improve the analytic model using Mindlin plate theory \cite{Mindlin51} which introduces two high frequency correction factors and therefore works for thick plates at higher frequency range. The explicit expressions for the transmission coefficient and the transmitted phase are obtained. Note that the analytic model for P-wave incidence in Ref.~\citen{Su2016} works very well, we will use this model in the discussion of the transmission properties for P-wave incidence. Several metasurface devices for different purposes, including mode splitting of SV- and P-waves and focusing of plane and cylindrical SV-waves, are designed using the analytic model and demonstrated using full wave FEM simulations.

The paper is organized as follows. The unit cell design and the transmission properties of the metasurface are introduced and discussed in Sec.~\ref{PhaseM}. The phase modulation by thickness variations of the plates is also proposed in Sec.~\ref{PhaseM}. Several metasurfaces are designed and demonstrated using full wave FEM simulations in Sec.~\ref{MSdesign}. Section \ref{concl} concludes the paper.

\section{\label{PhaseM}Unit cell design and transmission properties}
\subsection{Description of the transmission problem}
\begin{figure}[ht]
\centering
     \includegraphics[width=\columnwidth]{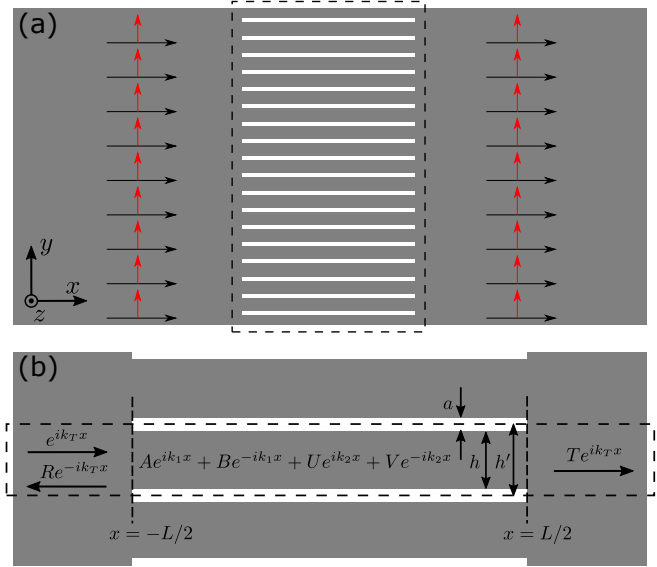}
\caption{2D schematic view; (a) shows two half-spaces connected by a uniform plate array separated by parallel cracks (white slits). The parallel black arrows indicate the propagation direction of the normally incident SV-wave; the red arrows indicate the direction of particle motion; (b) shows a unit cell of the slab.}\label{platearray}
\end{figure}

We first derive the transmission coefficient for a normally incident SV-wave  propagating through an array of uniform parallel plate separated by equally spaced rectangular cracks. The Young's modulus of the solid is denoted by $E$, shear modulus by $\mu$, Poisson's ratio by $\nu$ and density by $\rho$. The configuration is shown in Fig.~\ref{platearray}. 
Each individual plate has thickness $h$ and length $L$; the crack has width $a$ and length $L$. The transmission problem can be understood as follows: the normally incident plane SV-wave in a half-space impinges on an array of aligned parallel plates and couples with the flexural waves in plates; flexural waves then transmit through the plate array into another half-space producing the transmitted SV-wave. The critical  physics underlying the tranmission is that the phase speed of the SV-wave only depends on the material properties, while the flexural wave speed is sensitive to the thickness of each plate.  This combination of wave properties makes it possible to achieve specific phase delay through thickness variations. In order to quantify the metasurface design, it is useful to derive explicit expressions for the transmission coefficient and transmitted phase. Due to the periodicity in the $y$-direction, the whole transmission problem is equivalent to that  outlined in the dashed box in Fig.~\ref{platearray}(b). A detailed description regarding the notations in this figure will be given in Sec.~\ref{derive}.

The derivation procedure of the transmission and reflection coefficients for SV-wave incidence is similar to that in Ref.~\citen{Su2016} in which  Kirchhoff plate theory was used under thin plate assumptions. However, the Kirchhoff theory does not hold at the high frequency range even for thin plates. In this section, we will develop a more sophisticated model to better predict the transmission coefficient at the high frequency range. \citet{Mei2005} studied wave reflection and transmission in beams with discontinuities using Timoshenko beam theory. We are dealing with transverse wave reflection and transmission at the junction of bulk material and an array of plates where the governing equations are different. Here we use the Mindlin plate theory \cite{Mindlin51} with two high frequency correction factors and consider similar boundary conditions to establish the analytic model and calculate the transmission and reflection coefficients accurately. More detailed description and derivation will be given in the following sections. 

\subsection{Governing equations in the plate-like waveguides}\label{desc}
The plate-like waveguides connecting the two half-spaces act as phase modulators. 
We assume the plate array is uniform and due to the periodicity in the $y$-direction only consider the transmission problem in one plate, see Fig.~\ref{platearray}(b). According to the Mindlin plate theory, the governing equations of the transverse waves in the absence of external force   are
\begin{equation}\label{gov}
\begin{aligned}
\kappa \mu  \Big ( \frac{\partial \psi  }{\partial x} - \frac{\partial^2 w  }{\partial x^2} \Big )+\rho  \frac{\partial^2w }{\partial t^2}&=0,\\
EI\frac{\partial^2 \psi }{\partial x^2}+\kappa \mu A\Big( \frac{\partial w }{\partial x}-\psi  \Big)
-\lambda \rho I \frac{\partial^2 \psi }{\partial t^2} &= 0,
\end{aligned}
\end{equation}
where $w(x,t)$ is the displacement in the $y$-direction,  $\psi(x,t)$ is the  bending angle, 
$I=bh^3/12$ is the area moment of inertia; $A=bh$ is the cross section area of the plate; $\kappa$ and $\lambda$ are the shear and inertia correction factor, respectively. In the absence of external force, the free wave propagation solutions are   $w(x,t)=We^{i(kx-\omega t)}$ and  $\psi(x,t)=\Psi e^{i(kx-\omega t)}$. Plugging the solutions into Eq.~\eqref{gov} leads to
\begin{equation}\label{eig}
\begin{bmatrix}
\rho \omega^2 - \kappa \mu k^2   & -i\kappa \mu k\\
i\kappa \mu A k & \lambda \rho I \omega^2 -\kappa\mu A-EIk^2
\end{bmatrix}
\begin{pmatrix}
W\\ \Psi
\end{pmatrix}=0.
\end{equation}
The equation for the wavenumber $k$ is therefore 
\begin{equation}\label{dispersion}
k^4-(\lambda_P^2 +\frac{k^2_T}{\kappa})k^2+\frac{\lambda k^2_Tk_P^2}{\kappa}-k_F^4=0,
\end{equation}
where $k_T=\omega\sqrt{\rho/\mu}$, $k_P=\omega \sqrt{\rho(1-\nu^2)/E}$ and $k_F=\big( 12\omega^2\rho (1-\nu^2)/Eh^2 \big)^{1/4}$ are the wavenumbers of transverse, longitudinal and flexural waves, respectively. From Eq.~\eqref{dispersion}, the wavenumbers in the plate are
\begin{equation}\label{wavenum}
\begin{aligned}
k_1&=\pm \bigg\{ \frac{1}{2}(\lambda k_P^2+\frac{k^2_T}{\kappa})+\sqrt{\frac{1}{4}(\lambda k_P^2-\frac{k^2_T}{\kappa})^2+k^4_F} \bigg\}^{1/2},\\
k_2&=\pm \bigg\{ \frac{1}{2}(\lambda k_P^2+\frac{k^2_T}{\kappa})-\sqrt{\frac{1}{4}(\lambda k_P^2-\frac{k^2_T}{\kappa})^2+k^4_F} \bigg\}^{1/2}.
\end{aligned}
\end{equation}
Note that within the frequency range of interest ($\omega<\sqrt{12\mu\kappa/\lambda \rho h^2}$) $k_1$ is real and corresponds to a propagating wave, while $k_2$ is imaginary and corresponds to an evanescent wave.

\subsection{Transmission and reflection coefficients}\label{derive}
Considering the transmission problem for a normally incident SV-wave, in Fig.~\ref{platearray}(b), we assume the amplitude of the incident SV-wave as $1$, reflected SV-wave as $R$, transmitted SV-wave as $T$, and flexural waves in the plate array as $A_j$, $B_j$, $j=1,2$ so that
\begin{equation}\label{displ}
w= \begin{cases}
e^{ik_Tx}+Re^{-ik_Tx}, &x<- \frac L2,\\
A_1e^{ik_1x}+B_1e^{-ik_1x}  \\  \hspace{35pt} 
+A_2e^{ik_2x}+B_2e^{-ik_2x}, &|x|<\frac L2,\\
Te^{ik_Tx}, &x> \frac L2,
\end{cases}
\end{equation}
with time harmonic dependence $e^{-i\omega t}$ understood. The relations between the shear and inertia correction factors and the slope in the plate can be obtained from the equations of motion as \cite{Mindlin51}
\begin{equation}
\Psi=ik_j\beta_jW , \ \  \beta_j=1-\frac{k_T^2}{\kappa k_j^2}, \ \ j = 1 \text{ or }2.
\end{equation}
The deflection angles in different parts of the structure based on  Eq.~\eqref{displ} are 
\begin{equation}\label{slop}
\psi= \begin{cases}
ik_T\big(e^{ik_Tx}-Re^{-ik_Tx}\big), &x<-\frac L2,\\
ik_1\beta_1\big( A_1e^{ik_1x}-B_1e^{-ik_1x}\big)  \\  \hspace{10pt} 
+ik_2\beta_2\big( A_2e^{ik_2x}-B_2e^{-ik_2x}\big), & |x|<\frac L2,\\
ik_TTe^{ik_Tx}, &x>\frac L2,
\end{cases}
\end{equation}

Equations \eqref{displ} and \eqref{slop} involve six unknowns where the transmitted amplitude $T$ and the reflected amplitude $R$ are of particular interest. To solve for the six unknown parameters, we need six boundary conditions, i.e., displacement, deflection angle and shear force continuity at both ends of the plate. The displacement and deflection angle continuity can be easily established using Eqs.~\eqref{displ} and \eqref{slop}. The average shear forces along the $z$-direction in the half-space and the plate are $Q=\mu h^\prime(\partial w/\partial x)$ and $Q=\kappa \mu h(\partial w/\partial x -\psi)$, respectively. Using the six boundary conditions we can establish a $6 \times 6$ system with six unknowns to solve for the transmission and reflection coefficients. This  procedure involves too many undesired long equations, so as an alternative we   split the solutions into symmetric and antisymmetric modes, which reduces the problem  to two $3 \times 3$ systems. 

For the symmetric mode, the displacements in the half-spaces and the plate are rewritten as
\begin{equation}\label{symw}
w_S= \begin{cases}
\frac{1}{2}(e^{ik_Tx}+R_Se^{-ik_Tx}), &x<- \frac L2,\\
C_{S1}\cos k_1x  +C_{S2}\cos k_2x, &|x|< \frac L2 ,\\
\frac{1}{2}(e^{-ik_Tx}+R_Se^{ik_Tx}), &x>\frac L2 ,
\end{cases} 
\end{equation}
and the deflection angles are
\begin{equation}\label{sympsi}
\psi_S= \begin{cases}
\frac{ik_T}{2}(e^{ik_Tx}-R_Se^{-ik_Tx}), &x<- \frac L2,\\
-k_1\beta_1 C_{S1}\sin k_1x  -k_2\beta_2 C_{S2}\sin k_2x, &|x|< \frac L2 ,\\
-\frac{ik_T}{2}(  e^{-ik_Tx} - R_Se^{ik_Tx} ), &x>\frac L2 ,
\end{cases} 
\end{equation}
where $C_{S1}$ and $C_{S2}$ denote the amplitude of symmetric modes in the plate, $R_S$ corresponds to the amplitude of the reflected wave in the half-spaces. 
Similarly, considering the antisymmetric mode we can write the displacements and deflection angles as
\begin{equation}\label{antiw}
w_A= \begin{cases}
\frac{1}{2}(e^{ik_Tx}+R_Ae^{-ik_Tx}), &x<- \frac L2,\\
C_{A1}\sin k_1x  +C_{A2}\sin k_2x, & |x|< \frac L2 ,\\
-\frac{1}{2}(e^{-ik_Tx}+R_Ae^{ik_Tx}), &x>\frac L2 ,
\end{cases} 
\end{equation}
and 
\begin{equation}\label{antipsi}
\psi_A= \begin{cases}
\frac{ik_T}{2}(e^{ik_Tx}-R_Ae^{-ik_Tx}), &x<- \frac L2,\\
k_1\beta_1 C_{A1}\cos k_1x  +k_2\beta_2 C_{A2}\cos k_2x, &|x|< \frac L2 ,\\
\frac{ik_T}{2}( e^{-ik_Tx} - R_Ae^{ik_Tx} ), &x>\frac L2 ,
\end{cases} 
\end{equation}
where $C_{A1}$ and $C_{A2}$ denote the amplitude of antisymmetric modes in the plate, $R_A$ corresponds to the amplitude of the reflected wave in the half-spaces.

The boundary conditions for each problem now reduce to displacement, deflection angle and shear force continuity at one end of the plate. Applying the boundary conditions leads to 
\begin{widetext}
\begin{align}\label{sym}
\begin{bmatrix}
\cos(k_1L/2) & \cos(k_2L/2) & -\frac{1}{2}z\\
k_1\beta_1\sin(k_1L/2) & k_2\beta_2\sin(k_2L/2) & \frac{1}{2}ik_Tz\\
\kappa hk_1(1-\beta_1)\sin(k_1L/2) & \kappa hk_2(1-\beta_2)\sin(k_2L/2) & \frac{1}{2}ik_Th^\prime z
\end{bmatrix}
\begin{pmatrix}
C_{S1}\\ C_{S2}\\ R_S
\end{pmatrix}
&=
\begin{pmatrix}\frac{1}{2}z^{-1}\\ \frac{1}{2}ik_Tz^{-1}\\ \frac{1}{2}ik_Th^\prime z^{-1}\end{pmatrix} ,
\\  
\begin{bmatrix}
-\sin(k_1L/2) & -\sin(k_2L/2) & -\frac{1}{2}z\\
k_1\beta_1\cos(k_1L/2) & k_2\beta_2\cos(k_2L/2) & \frac{1}{2}ik_Tz\\
\kappa hk_1(1-\beta_1)\cos(k_1L/2) & \kappa hk_2(1-\beta_2)\cos(k_2L/2) & \frac{1}{2}ik_Th^\prime z
\end{bmatrix}
\begin{pmatrix}
C_{A1}\\ C_{A2}\\ R_A
\end{pmatrix} 
&=\begin{pmatrix}\frac{1}{2}z^{-1}\\ \frac{1}{2}ik_Tz^{-1}\\ \frac{1}{2}ik_Th^\prime z^{-1}\end{pmatrix},
\label{anti}
\end{align}
\end{widetext}
where $z= e^{ik_TL/2}$. 
Solving Eqs.~\eqref{sym} and \eqref{anti} yields the reflection coefficients for the symmetric and antisymmetric modes as 
\begin{equation}\label{rars}
\begin{aligned}
R_S&=
\frac{ \frac{\alpha_2}{k_1} \cot \frac{k_1L}2  + \frac{\alpha_1}{k_2 }  \cot \frac{k_2L}2
 +i\frac{(\beta_1-\beta_2)}{k_T} }
{ \frac{\alpha_2}{k_1} \cot \frac{k_1L}2  + \frac{\alpha_1}{k_2 }  \cot \frac{k_2L}2 
-i\frac{(\beta_1-\beta_2)}{k_T} }
e^{-ik_TL} ,\\
R_A&=
 \frac{  \frac{\alpha_2}{k_1} \cot \frac{k_1L}2  - \frac{\alpha_1}{k_2 }  \cot \frac{k_2L}2
+i\frac{(\beta_1-\beta_2)}{k_T} }
{  \frac{\alpha_2}{k_1} \cot \frac{k_1L}2  - \frac{\alpha_1}{k_2 }  \cot \frac{k_2L}2
-i\frac{(\beta_1-\beta_2)}{k_T} }
e^{-ik_TL} , 
\end{aligned}
\end{equation} 
where $\alpha_j = \beta_j -1 + \beta_j {h^\prime}/(\kappa h)$ for $j=1,2$. 
The transmission and reflection coefficients for the full problem are then 
\begin{equation}\label{transref}
\begin{aligned}
T&=\frac{1}{2}(R_S-R_A),\\
R&=\frac{1}{2}(R_S+R_A).
\end{aligned}
\end{equation}
The main result here is the transmission coefficient $T$ which not only shows the  amplitude but also contains information about the transmitted phase. 

\subsection{Numerical validation and phase modulation}
\begin{figure}[ht]
\centering
     \includegraphics[width=\columnwidth]{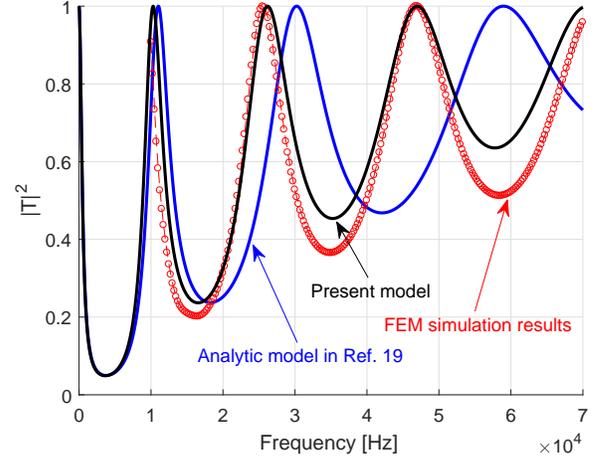} 
\caption{Comparison between analytical models and FEM simulation results. The blue curve corresponds to the frequency dependence of the transmission coefficient calculated using the model developed by \citet{Su2016}; the black curve corresponds to the transmission calculated using Eq.~\eqref{transref} in this paper; the red circles represent the FEM simulation results.}\label{compare}
\end{figure}

We now show that the theoretical model accurately predicts the transmission coefficient  using a numerical example. Consider an array of uniform plates in aluminum as described in Fig.~\ref{platearray}(a). The material properties are Young's modulus $E=70$ GPa, Poisson's ratio $\nu=0.33$ and density $\rho=2700$ kg/m$^3$. All the plates have  length $L=5$ cm and thickness $h=0.5$ cm and are separated by cracks of constant width $a=1$ mm. A plane SV-wave is normally incident from the left side of the metasurface. Figure \ref{compare} compares the present model and the theoretical model developed in Ref.~\citen{Su2016} to   FEM simulation results using COMSOL Multiphysics.
The Bloch-Floquet periodic condition was prescribed on the boundaries of the unit cell to mimic an infinite metasurface slab. The red curve in Fig.~\ref{compare} is calculated using the displacements extracted from the FEM simulation results. We can clearly see that the transmission curve calculated using the analytic model  agrees well with the FEM results at higher frequencies. It is also remarkable that the present model can accurately predict the total transmission frequencies, which correspond to the flexural resonances. This characteristic provides strong evidence that the predicted phase change  is close to the simulation results. Though the analytic model shows certain mismatch of amplitudes with the simulation results, the present model is still a useful tool since phase modulation is more crucial in metasurface design. This issue could potentially be addressed by using more complicated boundary conditions, \cite{Gregory1994} but is beyond the scope of this paper.

\begin{figure}[ht]
\centering
     \includegraphics[width=\columnwidth]{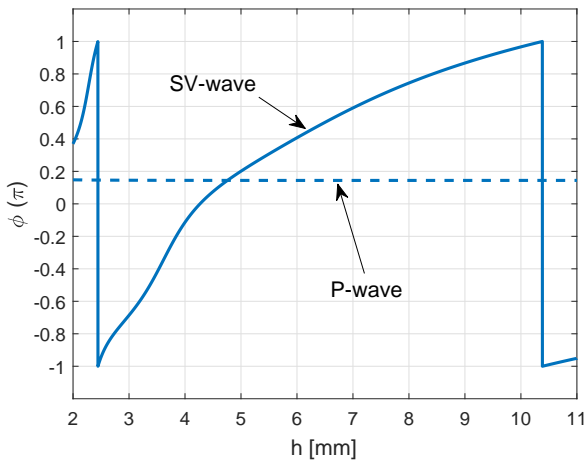}  
\caption{Transmitted phase at $60$ kHz through plates of different thicknesses. The solid line corresponds to the transmitted phase for a normally incident SV-wave; the dashed line is computed using the model developed by \citet{Su2016} and corresponds to the transmitted phase for P-wave incidence.}\label{phasec}
\end{figure}

The explicit expression for the transmitted phase can be easily extracted from Eq.~\eqref{transref} as
\begin{equation}\label{phasechange}
\phi=\tan^{-1} \big( {\text {imag}}(T)/{\text {real}}(T)\big ), \quad -\pi<\phi<\pi.
\end{equation}
According to the generalized Snell's law, i.e. Eq.~\eqref{general}, the metasurface design requires the transmitted wavefronts to cover the full $2\pi$ span. This can be easily satisfied using our design elements. The objective of this paper is to achieve full control of SV-wave using metasurface slabs with sub-wavelength thickness. In all the designs presented in Sec.~\ref{MSdesign}, the length of each plate is chosen as $L=5$ cm; the width of the gaps between plates is constant $a=1$ mm; the operation frequencies for all the metasurfaces are identical $60$ kHz, at which the wavelength of the transverse wave in the bulk material is larger than the slab thickness, i.e. $L<\lambda_T$. \rev{Though it is possible to design for a lower frequency range, $60$ kHz is selected to maintain relatively high transmission as shown in Fig.~\ref{compare}. } Consider a metasurface slab comprised of plates with different thickness in an aluminum background. The transmitted phases at $60$ kHz corresponding to different thicknesses of plates are calculated using Eq.~\eqref{phasechange} and plotted in Fig.~\ref{phasec}.
It is clear that the phase shifts through the internal plates, with thickness varying from $2.4$ mm to $10.4$ mm, cover a range of $2\pi$ for SV-wave incidence. \reva{Related to a point made earlier, all the plates are thinner than $11$ mm in the metasurface designs presented in this paper such that the minimum cutoff frequency for the $k_2$ mode is above $141.9$ kHz.}

The dashed line in Fig.~\ref{phasec} is computed using Eq.~(13) in Ref.~\citen{Su2016} and corresponds to the transmitted phase of a normally incident P-wave. It indicates  nearly identical  phase changes for different plate thickness. This can be  understood from the fact that the longitudinal wave speed in plates and P-wave speed in bulk material,
\begin{equation}\label{wavesp}
c_L=\sqrt{\frac{E}{\rho(1-\nu^2)}} \ \ {\text {and}} \ \
c_P=\sqrt{\frac{E(1-\nu)}{\rho(1+\nu)(1-2\nu)}},
\end{equation}
are both  functions of the material properties only. Phase modulation in metasurface designs is usually done by reducing the wave speed in each waveguide to achieve the desired phase delay. Thickness variation of the plate-like waveguides does not provide such a mechanism to modulate longitudinal waves in plates. The sppeds $c_L$ and $c_P$ are very close so that the effective P-wave impedance of the metasurface is similar to that of the background material. Therefore the normally incident P-wave will travel straight through the slab. Based on these properties, we can now design metasurface devices for various purposes such as mode splitting of SV- and P-waves and focusing of plane and cylindrical SV-waves.

\section{Applications in metasurface design}\label{MSdesign}
\subsection{\label{SplitSVP}Metasurface for splitting  SV- and P-waves}
\begin{figure}[ht]
\centering
     \includegraphics[width=0.95\columnwidth]{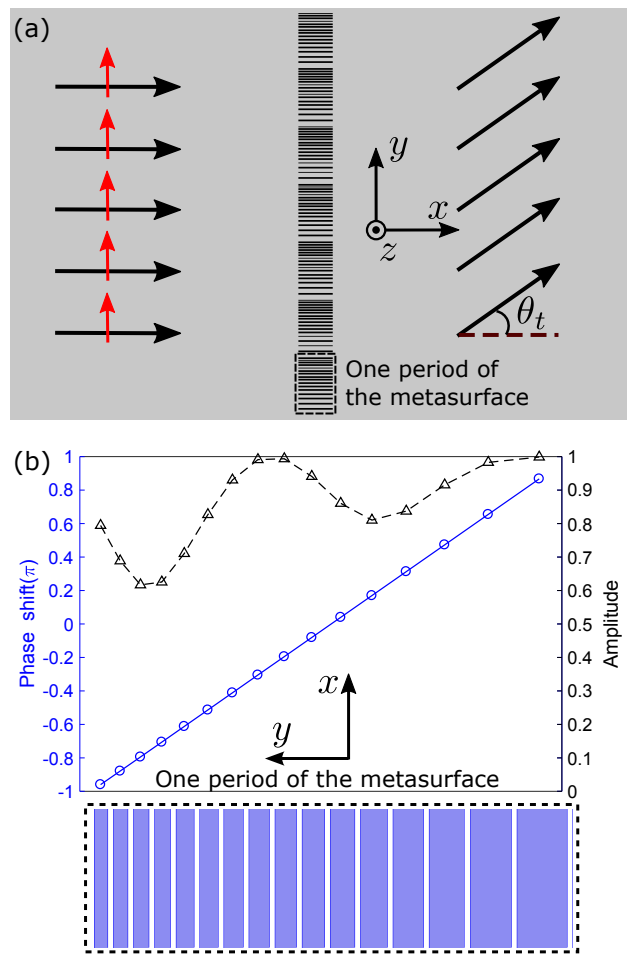}  
\caption{Metasurface design: (a)  outline of the design; (b)   transmission properties of each plate-like waveguide in one period of the metasurface. The blue circles and the black triangles represent the phase and amplitude of the transmitted wave through each plate, respectively.}\label{phaseamp}
\end{figure}

The material used in all the metasurface designs is aluminum with Young's modulus $E=70$ GPa, Poisson's ratio $\nu=0.33$ and density $\rho=2700$ kg/m$^3$. In the design for anomalous refraction of normally incident SV-wave, we choose a linear phase gradient $d\phi /dy=40\pi/\sqrt{3}$ rad/m. This phase gradient results in a transmitted angle of $\theta_t=30^\circ$ according to Eq.~\eqref{general}. The schematic view of the metasurface is illustrated in Fig.~\ref{phaseamp}(a). 
The width of the metasurface slab, i.e. length of each plate, is chosen as $L=5$ cm; the width of all the gaps are identical, $a=1$ mm. The thicknesses of the plates satisfying the constant phase gradient  are selected from Fig.~\ref{phasec}. Sixteen plates with thickness covering a complete phase change of $2\pi$ at $60$ kHz form one period of the metasurface as shown in the dashed box in Fig.~\ref{phaseamp}(a). The metasurface of infinite extent is formed by repeating the structure in the dashed box, such that the length of one period is $8.59$ cm.  The transmitted phases and amplitudes through all these plates are listed in Table \ref{tab}, while the 
\begin{table}[ht]\caption{Transmitted phase and amplitude through each L=$5$ cm long plate-like waveguide at $60$ kHz.} \label{tab}
\begin{ruledtabular}
\begin{tabular}{c c  c c}
 Plate  \ & $h$ (mm) & $\phi$ ($\pi$ rad) & $|T|$ (a.u.)  \\  
  \hline
   1  & 2.483 & -0.9599 & 0.7944\\ 
   2  & 2.586 & -0.8789 & 0.6886\\
   3  & 2.744 & -0.7946 & 0.6169\\ 
   4  & 2.953 & -0.7058 & 0.6253\\
   5  & 3.165 & -0.6121 & 0.7106\\ 
   6  & 3.351 & -0.5138 & 0.8268\\
   7  & 3.516 & -0.4113 & 0.9303\\ 
   8  & 3.676 & -0.3048 & 0.9901\\
   9  & 3.850 & -0.1950 & 0.9937\\ 
   10  & 4.069 & -0.0807 & 0.9410\\
   11  & 4.384 & 0.0398 & 0.8605\\ 
   12  & 4.864 & 0.1698 & 0.8103\\
   13  & 5.527 & 0.3129 & 0.8372\\ 
   14  & 6.345 & 0.4730 & 0.9154\\
   15  & 7.393 & 0.6547 & 0.9828\\ 
   16  & 9.000 & 0.8672 & 0.9988\\
\end{tabular}
\end{ruledtabular}
\end{table}
 transmitted amplitudes and phases in one period of the metasurface are plotted in Fig.~\ref{phaseamp}(b). Note that the plate array here is rotated $90^\circ$ counterclockwise, and this figure does not show the full length of each plate. It is clear that the transmitted phase strictly follows the constant spatial gradient. The amplitudes are not modulated in our design, but this does not affect the performance of the metasurface.

\begin{figure}[ht]
\centering
     \includegraphics[width=\columnwidth]{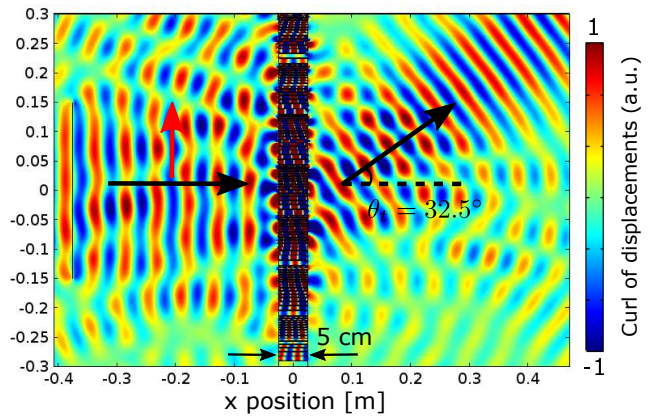}  
\caption{Anomalous refraction of a normally incident SV-wave at $60$ kHz. The curl of the displacement field is plotted to show the SV-wave.}\label{simu1}
\end{figure}

Full wave simulations were performed  in COMSOL to demonstrate the functionality of the metasurface at $60$ kHz. Displacements in the $y$-direction were applied along the vertical line at the left side of the simulation domain to generate an in-plane shear wave. Simulation result for SV-wave incidence is shown in Fig.~\ref{simu1}. 
The curl of the displacement field is plotted to show the distortional field of the wave, i.e. SV-wave. The black arrow indicates the propagation direction of the SV-wave, and the red arrow indicates the direction of particle motion of the incident wave. As indicated by the black arrow, a planar SV-wave is normally incident from the left side of the metasurface and is refracted at $32.5^\circ$ which is very close to the designed refraction angle. \rev{The metasurface can maintain this steering angle of the uniform transmitted beam over a wide frequency range from $55$ to $70$ kHz. The metasurface does not work  as effectively when the frequency is decreased or increased since the modulation is based on a single frequency.} It is worthwhile to point out that there is no mode converted wave, i.e. P-wave, in the far-field of the transmitted region. This can be explained as the modes in the transmitted field mainly come from the waveguides, i.e. plates. In this design, the SV-wave  impinges normally on the interface so that there is only a flexural wave in the plate array, which does not induce any mode converted wave in the transmitted field. It is noted that there are some longitudinal components in the interface wave at the right boundary of the metasurface, but they do not effect the far-field and hence do not influence the functionality of the metasurface. Near-field wave motion at the interface of the metasurface and the bulk material are not well studied and remain to be further investigated.

\begin{figure}[ht]
\centering
     \includegraphics[width=\columnwidth]{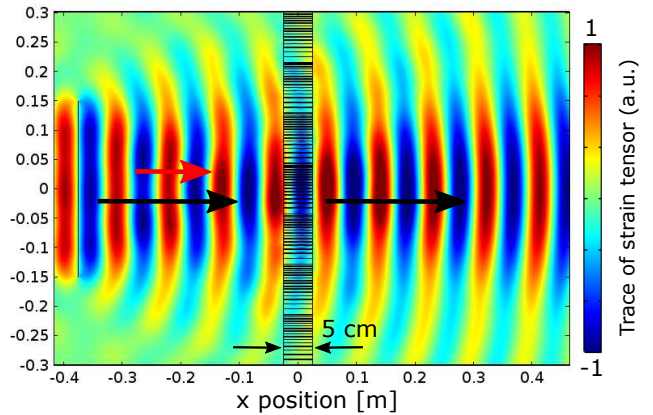}  
\caption{Unaffected normally incident P-wave at $60$ kHz. The trace of the strain tensor is plotted to show the P-wave.}\label{simu2}
\end{figure}

Another feature of the metasurface is that it does not alter the propagation direction of a normally incident P-wave;  it can therefore be used as a mode splitter to separate SV- and P-waves. Simulation results for normal incidence of P-wave are shown in Fig.~\ref{simu2}. In this case, displacements in the $x$-direction were applied along the vertical line at the left side of the simulation domain to generate a longitudinal wave, and 
 the trace of the strain tensor is plotted to show the dilatational field. The black arrow indicates the propagation direction of the P-wave, and the red arrow indicates the direction of particle motion of the incident wave. It is clear that the transmitted P-wave still travels along the normal direction. Similar to the SV-wave incidence case, there is no mode converted wave in the transmitted region. Since there are no mode conversions for both SV- and P-wave incidence, this metasurface is capable of steering these two types of waves into different directions without introducing unwanted wave types.

\subsection{\label{Focus}Metasurface for focusing  plane SV-waves}
Other than the metasurface for splitting SV- and P-waves, this approach can also be adopted in the design of a focal metasurface. The physics behind the focal metasurface is different from the gradient index lenses \cite{Climente2014,Tol2017,Su2017a,Su2017b} which are designed using ray theory and by varying the refractive indices of bulk materials. Here the focal metasurface is  based on the constructive and destructive interferences of the diffracted waves through all the waveguides. The required phase profile along the metasurface can be written as 
\begin{equation}\label{focpha}
\phi(y)=k_T\big( \sqrt{(y-y_0)^2+d^2} -d \big)+\phi_0, \ -\pi<\phi<\pi,
\end{equation}
where $y_0$ denotes the location where the metasurface is symmetric about and $\phi_0$ is the transmitted phase through the plate at $y_0$. 

\begin{figure}[ht]
\centering
     \includegraphics[width=\columnwidth]{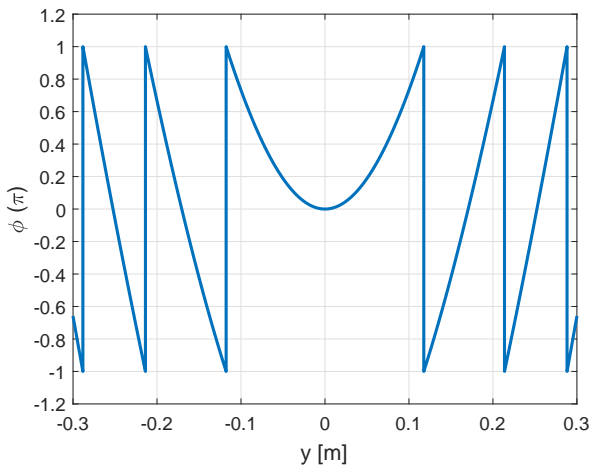}  
\caption{Phase profile of the metasurface for focusing a plane SV-wave at $60$ kHz.}\label{fph}
\end{figure}

We choose $y_0=0$ and set $\phi_0$=0 for convenience. The metasurface is chosen to have slab thickness, i.e. plate length, $L=5$ cm; the constant gap width is $a=1$ mm. The focal distance is selected to be $d=5L=25$ cm from the lens. The required phase profile at $60$ kHz is calculated using Eq.~\eqref{focpha} and plotted in Fig.~\ref{fph}.
A total number of $97$ plates are selected from Fig.~\ref{phasec} to form a metasurface. 

\begin{figure}[ht]
\centering
     \includegraphics[width=\columnwidth]{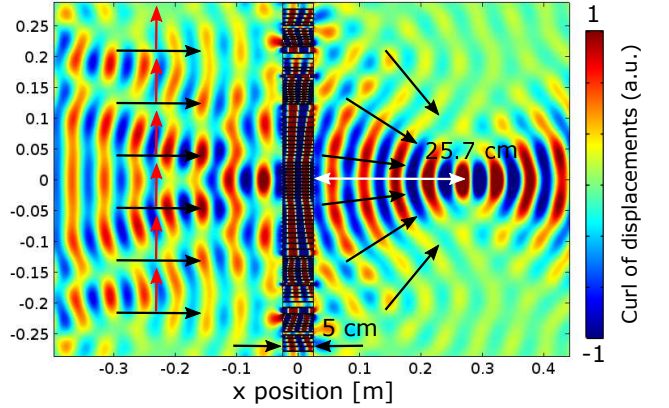}  
\caption{Focusing a normally incident plane SV-wave at $60$ kHz. The curl of the displacement field is plotted to show the SV-wave.}\label{simu3}
\end{figure}

Full wave FEM simulations using COMSOL demonstrate the focusing effect of the metasurface. Displacements in the $y$-direction were applied along the vertical line at the left side of the simulation domain to generate an in-plane shear wave. Figure \ref{simu3} shows the simulated field at $60$ kHz. 
The curl of the displacement field is plotted to represent the distortional wave. The focusing effect can be clearly observed at the right side of the metasurface. \rev{It is interesting that a focal spot can be observed over a wide range of frequency from $40$ to $75$ kHz, however, the focal distance is varying with frequency due to the dispersive nature of the design elements.} By comparing the energy density along the $x$-direction across the center of the metasurface, the focal point at $60$ kHz was found to be approximately $25.7$ cm away from the interface, which agrees with the designed distance to a remarkable degree.

\subsection{\label{Imag}Metasurface for focusing a cylindrical SV-wave}

We now design a metasurface for focusing a cylindrical SV-wave. Due to the cylindrical spreading of the wavefront, the phases of the incident wave along the metasurface is different, therefore Eq.~\eqref{focpha} needs to be revised. 
The  profile requires a  more rapid phase change as compared to the phase change for focusing a plane SV-wave.  
The modified phase profile can be written as \cite{Li2016}
\begin{equation}\label{imapha}
\phi(y)=k_T\big( \sqrt{(y-y_0)^2+d^2} -d \big)+\phi_c(y), \ -\pi<\phi<\pi,
\end{equation}
where $\phi_c(y)$ is a phase correction term which compensates the phase difference of the incident wave. 

\begin{figure}[ht]
\centering
     \includegraphics[width=\columnwidth]{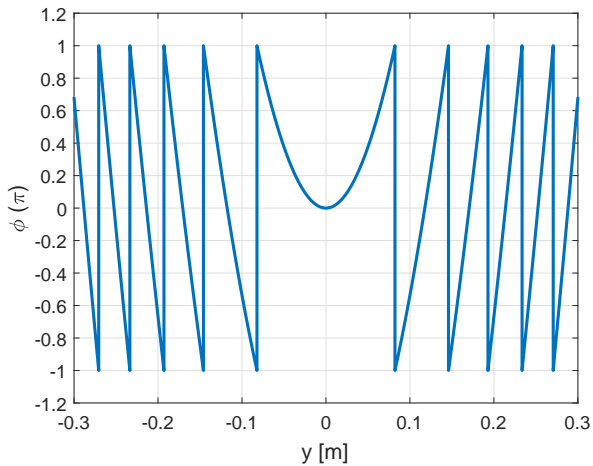}  
\caption{Phase profile of the metasurface for focusing a cylindrical SV-wave at $60$ kHz.}\label{ima}
\end{figure}

In this design the $y_0$ location, the slab thickness and the gap width are set the same as for the plane wave incidence design. The focal distance is selected to be $d=5L=25$ cm. The only difference is that the cylindrical wave source location, distance $d_S$ from the metasurface, needs to be taken into account. For instance, if we choose $d_S=d$ in this design, then the phase correction term is simply $\phi_c(y)=k_T\big( \sqrt{(y-y_0)^2+d^2} -d \big)$. The required phase profile at $60$ kHz is calculated using Eq.~\eqref{imapha} and plotted in Fig.~\ref{ima}.
Then a total number of $93$ plates are selected from Fig.~\ref{phasec} to form a metasurface.

\begin{figure}[ht]
\centering
     \includegraphics[width=\columnwidth]{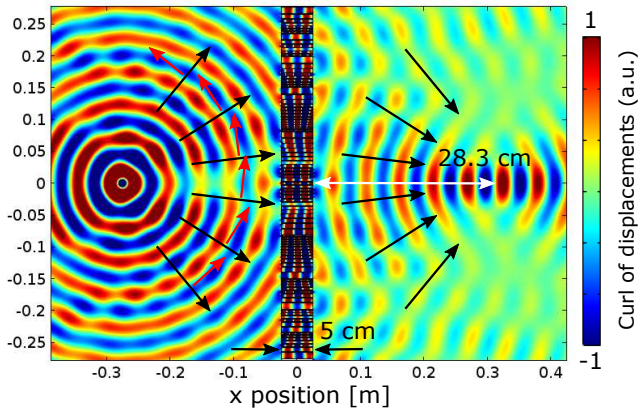}  
\caption{Focusing a cylindrical SV-wave source at $60$ kHz. The curl of the displacement field is plotted to show the SV-wave.}\label{simu4}
\end{figure}

The COMSOL-generated simulated SV-wave field at $60$ kHz is plotted in Fig.~\ref{simu4}. 
The curl field is shown to represent the distortional wave. The focal spot is evident at the right side of the metasurface. This suggests that our design elements are suitable for rapidly changing phase profiles. \rev{Similar to the previous focal metasurface for  a line source, this design also works over a broadband  frequency range from $45$ to $65$ kHz with focal distances varying with frequency.} \reva{It is also interesting to see the percentage of energy carried by the mode converted P-wave in the transmitted waves. Integration of the curl field and the strain were performed along the $y$-direction near the metasurface to estimate time-average of the power flux of the transmitted waves. Nearly $38.7\%$ of the transmitted energy is converted to P-waves.} The focal distance evaluated from the simulation results at $60$ kHz is $28.3$ cm, which is $3.3$ cm longer than the design distance. Given that the focal distance in the previous design is only $0.7$ cm longer than the designed distance, our model does not predict the transmitted phase accurately for oblique incidence. The main reason is that the current model does not consider the mode conversion which occurs for  oblique incidence. Improvement of the model to include such effects will be considered later. 

\reva{As a comparison, the COMSOL-generated simulated P-wave field at $60$ kHz is plotted in Fig.~\ref{pinci}. 
\setcounter{figure}{10}
\begin{figure}[hb!]
\centering
     \includegraphics[width=\columnwidth]{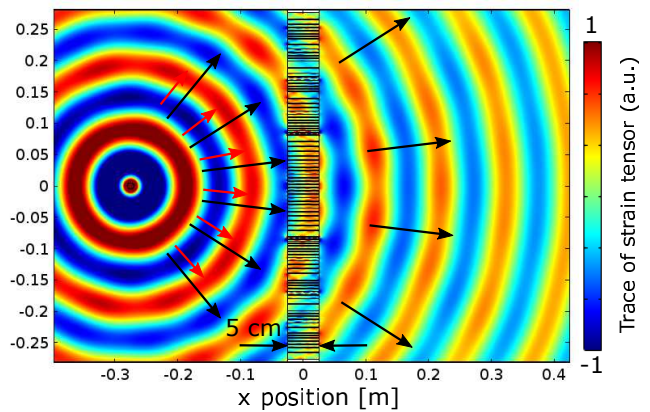}  
\caption{Cylindrical P-wave incidence at $60$ kHz. The trace of the strain tensor is plotted to show the P-wave.}\label{pinci}
\end{figure} 
The trace of the strain tensor is plotted to show the dilatational wave. It is clear that the phase change of the transmitted P-wave is almost negligible, and the wavefront is still cylindrical on the transmitted side. This is not surprising since the impedance of the slab is close to that of the background medium. Moreover, the slab width is sub-wavelength so that the phase shifts are small when the waves transmit through the plate array.}

\section{\label{concl}Conclusion}
In conclusion, we have presented a novel metasurface design approach for controlling SV-wave motion in elastic solids using plate-like waveguides of  varying thickness. A theoretical model based on the Mindlin plate theory is developed and compared with the FEM simulation results. The model works well for thick plates in the high frequency range and  is therefore well suted to metasurface design. It is also found that the transmission properties for normally incident P- and SV-waves are distinct. The transmitted phase of a normally incident SV-wave can cover a full span of $2\pi$. However, the transmitted phase of a normally incident P-wave is nearly constant. By taking advantage of these properties, we designed and numerically demonstrated several metasurfaces that are capable of steering SV-waves while remaining transparent to P-wave.  
The fundamental mode of this type of wave in a thick plate is nondispersive and can travel in a planar manner over long distances thus is of particular interest in nondestructive evaluations. \cite{Hirao1999,Lee2008,Petcher2014}

\section*{Acknowledgments}
This work was supported by Office of Naval Research through MURI Grant No.\ N00014-13-1-0631.

\section*{References}

\end{document}